\documentclass[pre,amsmath,amssymb,reprint]{revtex4-1}
\usepackage{epsfig}                                                                 
\usepackage{dcolumn}                                                                
\usepackage{bm}                                                                     

\usepackage[pdfstartview=FitH,bookmarksopen=true,bookmarksopenlevel=1]{hyperref}    
\begin{document}
\title[Nonexistence of two-dimensional sessile drops in the diffuse-interface model]%
{Nonexistence of two-dimensional sessile drops in the diffuse-interface model}
\author{E. S. Benilov}
 \altaffiliation[]{Department of Mathematics and Statistics, University of Limerick, Limerick, V94 T9PX, Ireland}
 \email{Eugene.Benilov@ul.ie}
 \homepage{https://staff.ul.ie/eugenebenilov/}
\date{\today}

\begin{abstract}
The diffuse-interface model (DIM) is a widely used tool for modeling fluid
phenomena involving interfaces -- such as, for example, sessile drops (liquid
drops on a solid substrate, surrounded by saturated vapor) and liquid ridges
(two-dimensional sessile drops). In this work, it is proved that,
surprisingly, the DIM does not admit solutions describing static liquid
ridges. If, however, the vapor-to-liquid density ratio is small -- as, for
example, for water at room temperature -- the ridges can still be observed as
quasi-static states, as their evolution is too slow to be distinguishable from
evaporation. Interestingly, the nonexistence theorem cannot be extended to
axisymmetric sessile drops and ridges near a vertical wall, which are not
ruled out.
\end{abstract}
\maketitle

\section{Introduction}

The diffuse-interface model (DIM) was developed in Refs.
\cite{HohenbergHalperin77,Gouin87,AndersonMcFaddenWheeler98,PismenPomeau00,Jacqmin00,ThieleMadrugaFrastia07}
and used for modeling various interfacial phenomena (e.g., Refs.
\cite{DingSpelt07,YueFeng11,SibleyNoldSavvaKalliadasis13b,KusumaatmajaHemingwayFielding16,FakhariBolster17,BorciaBorciaBestehornVarlamovaHoefnerReif19,GalloMagalettiCoccoCasciola20,GelissenVandergeldBaltussenKuerten20,Benilov20a,Benilov20b}
and references therein). Unlike phenomenological models, the DIM is based on a
physical assumption: that the van der Waals intermolecular force is described
by a pair-wise potential whose spatial scale is much smaller than that of the flow.

In the present work, the DIM is applied to a static liquid--vapor interface
located near a flat horizontal substrate. The fluid between the substrate and
interface is in liquid phase, whereas the fluid above the interface is vapor.
In three dimensions, such a configuration is usually referred to as a
\textquotedblleft sessile drop\textquotedblright, and in two dimensions, a
\textquotedblleft liquid ridge\textquotedblright\ (hereinafter, just
\textquotedblleft ridge\textquotedblright).

The main conclusion of the present work will come as a surprise: the DIM does
\emph{not} admit solutions describing static ridges. The proof of the
nonexistence theorem will be presented for the van der Waals fluid (Theorem
1), but it can be readily extended to any equation of state (Theorem 2). This
result seems to contradict the existence of ridge solutions in the
Navier--Stokes equations, but the apparent conflict will be convincingly
resolved.\smallskip

\section{Formulation}

Let a fluid's density field be
two-dimensional and described by a function $\rho(x,z)$ where $x$ and $z$ are
the horizontal and vertical coordinates, respectively. The following
nondimensional variables will be used:%
\[
\rho_{nd}=b\rho,\qquad x_{nd}=\left(  \frac{a}{K}\right)  ^{1/2}x,\qquad
z_{nd}=\left(  \frac{a}{K}\right)  ^{1/2}z,
\]
where $a$ and $b$ are the van der Waals constants ($b^{-1}$ is the maximum
density), and $K$ is the Korteweg parameter (characterizing the fluid--fluid
intermolecular force). A nondimensional temperature can be defined by%
\[
\tau=\frac{RTb}{a},
\]
where $R$ is the gas constant and $T$, the dimensional temperature.

In the equilibrium case (zero flow, constant temperature), the DIM reduces to
a single equation for the nondimensional density field (e.g., Eq. (34) of Ref.
\cite{Benilov20a}). For the van der Waals fluid, this equation is (the
subscript $_{nd}$ omitted)%
\begin{equation}
\tau\left(  \ln\frac{\rho}{1-\rho}+\frac{1}{1-\rho}\right)  -2\rho
-\frac{\partial^{2}\rho}{\partial x^{2}}-\frac{\partial^{2}\rho}{\partial
z^{2}}=\mu, \label{1}%
\end{equation}
where $\mu$ is an undetermined constant. Note that, physically, the first two
terms on the left-hand side of Eq. (\ref{1}) represent the chemical potential
of the van der Waals fluid.

Let the fluid be bounded below by a solid substrate located at $z=0$, in which
case the DIM implies the following boundary condition
\cite{PismenPomeau00,Benilov20a}:%
\begin{equation}
\rho=\rho_{0}\qquad\text{at}\qquad z=0, \label{2}%
\end{equation}
where $\rho_{0}$ characterizes the solid--fluid intermolecular force. Far
above the substrate, $\rho$ tends to the density $\rho_{v}$ of saturated
vapor,%
\begin{equation}
\rho\rightarrow\rho_{v}\qquad\text{as}\qquad z\rightarrow\infty. \label{3}%
\end{equation}
The vapor density $\rho_{v}$ and the matching liquid density $\rho_{l}$ are
determined by the so-called Maxwell construction, comprising the requirements
that the vapor's pressure and chemical potential be equal to those of the
liquid (e.g., Ref. \cite{Benilov20a}). For the van der Waals fluid, the
Maxwell construction is%
\begin{equation}
\frac{\tau\rho_{v}}{1-\rho_{v}}-\rho_{v}^{2}=\frac{\tau\rho_{l}}{1-\rho_{l}%
}-\rho_{l}^{2}, \label{4}%
\end{equation}%
\begin{multline}
\tau\left(  \ln\frac{\rho_{v}}{1-\rho_{v}}+\frac{1}{1-\rho_{v}}\right)
-2\rho_{v}\\
=\tau\left(  \ln\frac{\rho_{l}}{1-\rho_{l}}+\frac{1}{1-\rho_{l}}\right)
-2\rho_{l}. \label{5}%
\end{multline}
Unlike $\rho_{v}$, the liquid density $\rho_{l}$ is not involved in the
boundary-value problem for $\rho(x,z)$.

\begin{figure}
\includegraphics[width=\columnwidth]{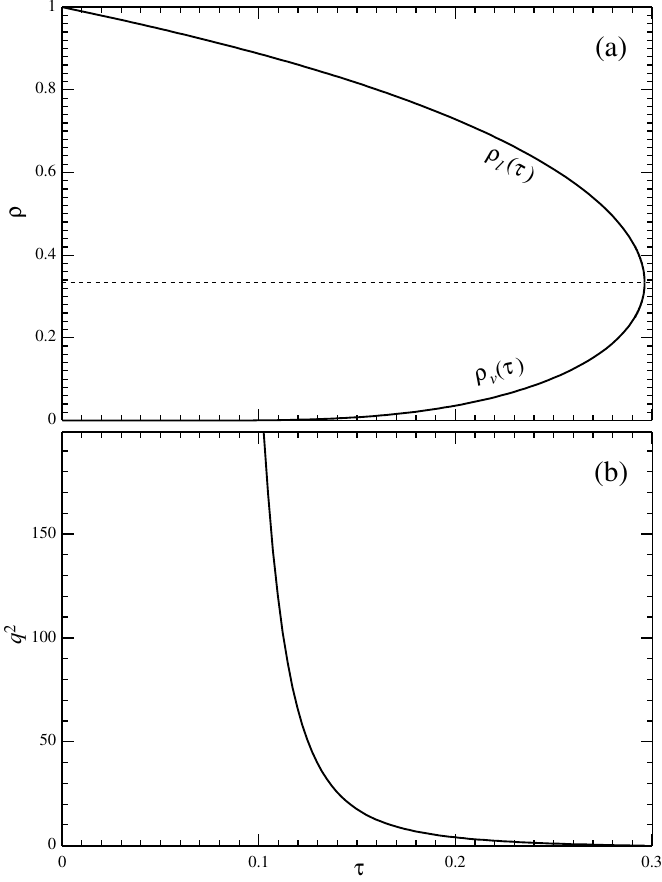}
\caption{The solution of the Maxwell construction (\ref{4})-(\ref{5}) for the van der Waals fluid vs. the nondimensional temperature: (a) the densities of saturated vapor and liquid (the dotted line shows the critical density), (b) $q^{2}$ given by expression (\ref{14}).}
\label{fig1}
\end{figure}

Eqs. (\ref{4})-(\ref{5}) admit non-trivial ($\rho_{v}\neq\rho_{l}$) solutions
only if the temperature is lower than its critical value. Nondimensionally,
this restriction amounts to $\tau<8/27$ and is implied everywhere in this
paper (otherwise interfaces simply do not exist). The graphs of $\rho_{v}$ and
$\rho_{l}$ vs. $\tau$ are shown in Fig. \ref{fig1}a.

In principle, the boundary condition (\ref{3}) can be replaced with%
\[
\rho\rightarrow\rho_{\infty}\qquad\text{as}\qquad z\rightarrow\infty,
\]
where $\rho_{\infty}$ is not necessarily equal to $\rho_{v}$. What happens in
this case depends on whether $\rho_{\infty}$ exceeds $\rho_{v}$. If it does,
the vapor above the ridge is over-saturated -- hence, unstable with respect to
spontaneous formation of drops. Even if static solutions exist in this case,
they are meaningless physically.

If, on the other hand, the vapor is under-saturated ($\rho_{\infty}<\rho_{v}%
$), ridges cannot exist due to evaporation. This conclusion is based on
physics, but can be readily confirmed mathematically, as this case is similar
to the one actually examined ($\rho_{\infty}=\rho_{v}$).

According to the DIM, ridges are accompanied by a precursor film stretching to
infinity \cite{PismenPomeau00}, so that%
\begin{equation}
\rho\rightarrow\bar{\rho}(z)\qquad\text{as}\qquad x\rightarrow\infty,
\label{6}%
\end{equation}
where $\bar{\rho}(z)$ describes the vertical structure of the precursor film
and satisfies the one-dimensional reduction of the boundary-value problem
(\ref{1})-(\ref{3}),%
\begin{equation}
\tau\left(  \ln\frac{\bar{\rho}}{1-\bar{\rho}}+\frac{1}{1-\bar{\rho}}\right)
-2\bar{\rho}-\frac{\mathrm{d}^{2}\bar{\rho}}{\mathrm{d}z^{2}}-\mu=0, \label{7}%
\end{equation}%
\begin{equation}
\bar{\rho}=\rho_{0}\qquad\text{at}\qquad z=0, \label{8}%
\end{equation}%
\begin{equation}
\bar{\rho}\rightarrow\rho_{v}\qquad\text{as}\qquad z\rightarrow\infty.
\label{9}%
\end{equation}
One can safely assume that, if exists, the ridge solution is symmetric -- say,
with respect to $x=0$ -- hence,%
\begin{equation}
\frac{\partial\rho}{\partial x}=0\qquad\text{at}\qquad x=0. \label{10}%
\end{equation}
To eliminate the trivial solution (describing the precursor film without a
ridge), require%
\begin{equation}
\rho\neq\bar{\rho}\qquad\text{at}\qquad x=0. \label{11}%
\end{equation}
This condition is sufficient if it holds in any non-zero-length interval of
$z$, no matter how small. Finally, let%
\begin{equation}
\left\vert \int_{-\infty}^{\infty}\left(  \rho-\bar{\rho}\right)
\mathrm{d}x\right\vert <\infty\qquad\forall z>0. \label{12}%
\end{equation}
This condition eliminates solutions describing infinite arrays of ridges (as
opposed to a single, isolated one).\smallskip

\section{Properties of the precursor film}

The following properties of the function $\bar{\rho}(z)$ will be needed.\smallskip

(i) Taking the limit $z\rightarrow\infty$ in Eq. (\ref{1}) and recalling the
boundary condition (\ref{3}), one obtains%
\begin{equation}
\mu=\tau\left(  \ln\frac{\rho_{v}}{1-\rho_{v}}+\frac{1}{1-\rho_{v}}\right)
-2\rho_{v}. \label{13}%
\end{equation}
Now, multiply Eq. (\ref{7}) by $\mathrm{d}\bar{\rho}/\mathrm{d}z$ and
integrate it with respect to $z$, which yields a separable first-order
equation. Using the boundary condition (\ref{9}) to fix the constant of
integration in this equation and taking into account (\ref{13}), one can
extend $\bar{\rho}(z)$ to $z<0$, then show that it is monotonic, and
eventually use the Maxwell construction (\ref{4})-(\ref{5}) to prove that%
\[
\bar{\rho}(z)\rightarrow\rho_{l}\qquad\text{as}\qquad z\rightarrow-\infty.
\]
The extended $\bar{\rho}(z)$ describes a liquid--vapor interface in an
unbounded space. Given its monotonicity, the boundary condition (\ref{8}) may
hold only if $\rho_{v}<\rho_{0}<\rho_{l}$. This requirement is implied
everywhere in this paper.\smallskip

(ii) Using the above-mentioned separable equation, one can show that the
solution of the boundary-value problem (\ref{7})-(\ref{9}) for $\bar{\rho}(z)$
is unique.\smallskip

(iii) Introduce%
\begin{equation}
q^{2}=\frac{\tau}{\rho_{v}\left(  1-\rho_{v}\right)  ^{2}}-2. \label{14}%
\end{equation}
It can be deduced from the Maxwell construction (\ref{4})-(\ref{5}) that%
\begin{equation}
q^{2}>0\qquad\text{if}\qquad\tau<\frac{8}{27} \label{15}%
\end{equation}
(see also Fig. \ref{fig1}b). Then, it follows from (\ref{7}), (\ref{13}), and
(\ref{9}) that%
\begin{equation}
\bar{\rho}=\rho_{v}+\mathcal{O(}\operatorname{e}^{-qz})\qquad\text{as}\qquad
z\rightarrow\infty, \label{16}%
\end{equation}
where it is implied that $q>0$.\smallskip

\section{The nonexistence theorem}

The main result of the present work can be formulated as follows.\smallskip

\noindent\textbf{Theorem \textbf{1}:} \emph{The boundary-value problem
(\ref{1})-(\ref{12}) does not admit smooth solutions for }$\rho(x,z)$%
\emph{.}\smallskip

Theorem 1 follows from the four lemmas formulated below. Lemmas 1-2 prove
that, if a solution exists, it should satisfy a certain identity, whereas
Lemmas 3-4 prove that the right-hand side of this identity is strictly smaller
than its left-hand side. The resulting contradiction can only be resolved if
Eqs. (\ref{1})-(\ref{12}) do not admit any solutions.\smallskip

\noindent\textbf{Lemma 1:} \emph{All solutions of Eqs. (\ref{1})-(\ref{12})
(if exist) satisfy}%
\begin{equation}
F[\rho(0,z)]=F[\bar{\rho}(z)], \label{17}%
\end{equation}
\emph{where the functional }$F[\varrho(z)]$\emph{ is given by}%
\begin{multline*}
F[\varrho(z)]=\int_{0}^{\infty}\left[  \tau\varrho\ln\frac{\varrho}{1-\varrho
}-\varrho^{2}%
\vphantom{\left( \frac{\mathrm{d}\varrho}{\mathrm{d}\varrho }\right) ^{2}}\right.
\\
+\left.  \frac{1}{2}\left(  \frac{\mathrm{d}\varrho}{\mathrm{d}z}\right)
^{2}-\mu\varrho+p\right]  \mathrm{d}z
\end{multline*}
\emph{and}{%
\[
p=\frac{\tau\rho_{v}}{1-\rho_{v}}-\rho_{v}^{2}.
\]
Note that, physically, }$p$ is the vapor pressure far above the ridge.

To prove Lemma 1, multiply Eq. (\ref{1}) by $\partial\rho/\partial x$ and
integrate it with respect to $z$ from $0$ to $\infty$. Integrating the term
involving $\partial^{2}\rho/\partial z^{2}$ by parts, and using the boundary
conditions (\ref{2})-(\ref{3}), one obtains%
\begin{equation}
\int_{0}^{\infty}\frac{\partial f}{\partial x}\mathrm{d}z=0, \label{18}%
\end{equation}
where%
\begin{equation}
f=\tau\rho\ln\frac{\rho}{1-\rho}-\rho^{2}-\frac{1}{2}\left(  \frac
{\partial\rho}{\partial x}\right)  ^{2}+\frac{1}{2}\left(  \frac{\partial\rho
}{\partial z}\right)  ^{2}-\mu\rho. \label{19}%
\end{equation}
Using the boundary condition (\ref{3}) and expression (\ref{13}) for $\mu$,
one can show that $f\rightarrow-p$ as $z\rightarrow\infty$, with $f+p$
vanishing exponentially quickly (as follows from Lemma 2 proved below). This
enables one to rewrite (\ref{18}) in the form%
\[
\frac{\partial}{\partial x}\int_{0}^{\infty}\left(  f+p\right)  \mathrm{d}%
z=0.
\]
To obtain identity (\ref{17}), one should integrate the above equality with
respect to $x$ from $0$ to $\infty$, replace $f$ with expression (\ref{19}),
and use conditions (\ref{6}) and (\ref{10}).\smallskip

\noindent\textbf{Lemma 2:} \emph{Solutions of Eqs. (\ref{1})-(\ref{12}) (if
exist) are such that}%
\begin{multline}
\rho(x,z)\sim\rho_{v}\\
+\frac{C}{z^{1/2}}\exp\left(  -qz-\frac{qx^{2}}{2z}\right)  \qquad
\text{as}\qquad z\rightarrow\infty, \label{20}%
\end{multline}
\emph{where }$q>0$\emph{ is given by (\ref{14}) and }$C$\emph{ is a
constant.}\smallskip

Let $\rho=\bar{\rho}+\tilde{\rho}$. Under the assumption that $\rho$ exists
and satisfies Eq. (\ref{1}), the large-$z$ asymptotics of $\tilde{\rho}(x,z)$
should vanish as $z\rightarrow\infty$ and satisfy the linearized version of
Eq. (\ref{1}),%
\begin{equation}
\tilde{\rho}\rightarrow0\qquad\text{as}\qquad z\rightarrow\infty, \label{21}%
\end{equation}%
\begin{equation}
q^{2}\tilde{\rho}-\frac{\partial^{2}\tilde{\rho}}{\partial x^{2}}%
-\frac{\partial^{2}\tilde{\rho}}{\partial z^{2}}=0\qquad\text{as}\qquad
z\rightarrow\infty. \label{22}%
\end{equation}
The general solution of Eq. (\ref{22}) subject to condition (\ref{21}) can be
found via the Fourier transformation with respect to $x$, which yields%
\begin{multline*}
\tilde{\rho}\sim\int_{-\infty}^{\infty}B(k)\\
\times\exp\!\left(  -z\sqrt{q^{2}+k^{2}}+\mathrm{i}xk\right)  \mathrm{d}%
k\qquad\text{as}\qquad z\rightarrow\infty,
\end{multline*}
where $B(k)$ is an undetermined function. By virtue of (\ref{12}), $B(k)$ is
continuous for all $k$ including $k=0$ -- on the basis of which the above
integral can be simplified and eventually evaluated,%
\[
\tilde{\rho}\sim B(0)\sqrt{\frac{2\pi q}{z}}\exp\left(  -zq-\frac{qx^{2}}%
{2z}\right)  \qquad\text{as}\qquad z\rightarrow\infty.
\]
Recalling the definition of $\tilde{\rho}$ and letting $B(0)\sqrt{2\pi q}=C$,
one can transform the above expression into (\ref{20}), as required.\smallskip

\noindent\textbf{Lemma 3:} \emph{The function }$\bar{\rho}(z)$\emph{ minimizes
the functional }$F[\varrho(z)]$\emph{ \emph{under additional constraints}}%
\begin{equation}
\varrho=\rho_{0}\qquad\text{at}\qquad z=0, \label{23}%
\end{equation}%
\begin{equation}
\varrho\rightarrow\rho_{v}\qquad\text{as}\qquad z\rightarrow\infty. \label{24}%
\end{equation}
To understand the role of Lemma 3 in proving Theorem 1, note that the former
and condition (\ref{11}) make it impossible for identity (\ref{17}) to hold --
thus creating the desired contradiction.

Let $\delta\varrho(z)$ be a variation of $\varrho(z)$, so that the boundary
conditions (\ref{23})-(\ref{24}) yield%
\begin{align*}
\delta\varrho &  =0\qquad\text{at}\qquad z=0,\\
\delta\varrho &  \rightarrow0\qquad\text{as}\qquad z\rightarrow\infty.
\end{align*}
Under these conditions, the requirement%
\[
\delta F[\varrho(z)]=0
\]
yields an equation for $\varrho$, coinciding with Eq. (\ref{7}) for $\bar
{\rho}$. Conditions (\ref{23})-(\ref{24}), in turn, coincide with the
conditions (\ref{8})-(\ref{9}) for $\bar{\rho}$. Thus, $\varrho=\bar{\rho}$ is
a stationary point of $F[\varrho(z)]$.

Recall also that $\bar{\rho}(z)$ is unique -- hence, $F[\varrho(z)]$ has only
\emph{one} stationary point. If it happens to be a minimum, it is the
\emph{absolute} minimum of the functional in question.

To ensure that $\varrho=\bar{\rho}$ is indeed a minimum of $F[\varrho(z)]$,
one has to prove that%
\begin{equation}
\delta^{2}F[\varrho(z)]>0\qquad\text{for}\qquad\varrho=\bar{\rho}. \label{25}%
\end{equation}
To do so, observe that%
\[
\delta^{2}F[\varrho(z)]=\int_{0}^{\infty}\delta\varrho\,\hat{A}\,\delta
\varrho\,\mathrm{d}z,
\]
where%
\[
\hat{A}=\frac{\tau}{\bar{\rho}\left(  1-\bar{\rho}\right)  ^{2}}%
-2-\frac{\mathrm{d}^{2}}{\mathrm{d}z^{2}}%
\]
is a second-order self-adjoint non-singular operator -- hence, its spectrum is
real and the eigenfunctions form an orthogonal basis in $L^{2}(0,\infty)$
\cite{CourantHilbert89}. Then, condition (\ref{25}) holds if and only if
$\hat{A}$ is positive-definite -- which it indeed is, as follows from Lemma
4.\smallskip

\noindent\textbf{Lemma 4:} \emph{All eigenvalues of }$\hat{A}$\emph{ are
strictly positive.}\smallskip

Let $\lambda$ be a \emph{continuous}-spectrum eigenvalue of $\hat{A}$, and
$\psi$ the corresponding eigenfunction,
\begin{equation}
\left[  \frac{\tau}{\bar{\rho}\left(  1-\bar{\rho}\right)  ^{2}}-2\right]
\psi-\frac{\mathrm{d}^{2}\psi}{\mathrm{d}z^{2}}=\lambda\psi, \label{26}%
\end{equation}%
\begin{equation}
\psi=0\qquad\text{at}\qquad z=0, \label{27}%
\end{equation}%
\begin{equation}
\psi\sim\cos(\kappa_{c}z+\theta_{c})\qquad\text{as}\qquad z\rightarrow\infty,
\label{28}%
\end{equation}
where $\theta_{c}$ is an undetermined constant, $\kappa_{c}=\sqrt
{\lambda-q^{2}}$ with $q^{2}$ given by (\ref{14}), and it is implied that
$\kappa_{c}$ is real -- hence,%
\[
\lambda\geq q^{2}.
\]
Given (\ref{15}), the above inequality entails $\lambda>0$ as required.

To examine the \emph{discrete} spectrum (if it exists), assume that
$\lambda<q^{2}$ and replace (\ref{28}) with
\begin{equation}
\psi=\mathcal{O}(\operatorname{e}^{-\kappa_{d}z})\qquad\text{as}\qquad
z\rightarrow\infty, \label{29}%
\end{equation}
where $\kappa_{d}=\sqrt{q^{2}-\lambda}$. Next, introduce $\phi(z)$ such that%
\begin{equation}
\psi=\frac{\mathrm{d}\bar{\rho}}{\mathrm{d}z}\phi. \label{30}%
\end{equation}
Substituting this expression into (\ref{26})-(\ref{27}) and (\ref{29}), and
recalling that $\bar{\rho}$ satisfies (\ref{7}) and (\ref{16}), one obtains%
\begin{equation}
-\frac{\mathrm{d}}{\mathrm{d}z}\left[  \left(  \frac{\mathrm{d}\bar{\rho}%
}{\mathrm{d}z}\right)  ^{2}\frac{\mathrm{d}\phi}{\mathrm{d}z}\right]
=\lambda\left(  \frac{\mathrm{d}\bar{\rho}}{\mathrm{d}z}\right)  ^{2}\phi,
\label{31}%
\end{equation}%
\begin{equation}
\phi=0\qquad\text{at}\qquad z=0, \label{32}%
\end{equation}%
\begin{equation}
\phi=\mathcal{O}(\operatorname{e}^{-\kappa_{d}z+qz})\qquad\text{as}\qquad
z\rightarrow\infty\label{33}%
\end{equation}
Now, multiply Eq. (\ref{31}) by $\phi$ and integrate it with respect to $z$
from $0$ to $\infty$. Integrating by parts and taking into account conditions
(\ref{32})-(\ref{33}) and (\ref{16}), one obtains%
\[
\int_{0}^{\infty}\left(  \frac{\mathrm{d}\varrho}{\mathrm{d}z}\right)
^{2}\left(  \frac{\mathrm{d}\phi}{\mathrm{d}z}\right)  ^{2}\mathrm{d}%
z=\lambda\int_{0}^{\infty}\left(  \frac{\mathrm{d}\varrho}{\mathrm{d}%
z}\right)  ^{2}\phi^{2}\mathrm{d}z.
\]
Both integrals in the above identity are strictly positive and converge (the
latter, because their integrands decay exponentially as $z\rightarrow\infty$).
Hence, $\lambda>0$, as required.

The proof of Theorem 1 is now complete. Next, it will be extended to an
arbitrary equation of state (not necessarily the van der Waals one).\smallskip

\noindent\textbf{Theorem 2:} \emph{Let smooth functions }$G(\rho,\tau)$\emph{
and }$p(\rho,\tau)$\emph{ be such that }$\partial p/\partial\rho
=\rho\,\partial G/\partial\rho$\emph{, and consider the \emph{following
}extension of Eq. (\ref{1}):}%
\begin{equation}
G(\tau,\rho)-\frac{\partial^{2}\rho}{\partial x^{2}}-\frac{\partial^{2}\rho
}{\partial z^{2}}=\mu,\label{34}%
\end{equation}
\emph{that of the Maxwell construction (\ref{4})-(\ref{5}):}%
\begin{equation}
G(\rho_{v},\tau)=G(\rho_{l},\tau),\qquad p(\rho_{v},\tau)=p(\rho_{l}%
,\tau),\label{35}%
\end{equation}
\emph{and that of the precursor film problem (\ref{7})-(\ref{9}):}%
\begin{equation}
G(\tau,\bar{\rho})-\frac{\mathrm{d}^{2}\bar{\rho}}{\mathrm{d}z^{2}}%
-\mu=0,\label{36}%
\end{equation}%
\begin{equation}
\bar{\rho}=\rho_{0}\qquad\text{at}\qquad z=0,\label{37}%
\end{equation}%
\begin{equation}
\bar{\rho}\rightarrow\rho_{v}\qquad\text{as}\qquad z\rightarrow\infty
.\label{38}%
\end{equation}
\emph{(\ref{34})-(\ref{38}) and the (old) boundary conditions (\ref{2}%
)-(\ref{3}), (\ref{6}), (\ref{10})-(\ref{12}), do not admit smooth
solutions\emph{ describing liquid ridges}.}\smallskip

The proof of Theorem 2 will not be presented, as it is similar to that of
Theorem 1, e.g., the van der Waals expression (\ref{14}) should be replaced
with%
\[
q^{2}=\left(  \frac{\partial G}{\partial\rho}\right)  _{\rho=\rho_{v}},
\]
etc. Thus, the DIM does not admit solutions for an isolated static ridge in
\emph{any} fluid.\smallskip

\section{Discussion}

The most counter-intuitive aspect of Theorems
1-2 is that it rules out a phenomenon which one can easily reproduce by
depositing a streak of water on one's kitchen table. There seems to be a
mathematical paradox too: ridges do exist in the Navier--Stokes equations,
which can be viewed as a liquid-only, incompressible reduction of the DIM.

The resolution of the paradox is best illustrated under the assumption that
the interfacial slope $\varepsilon$ is small, i.e., in the thin-film
approximation. The study of this limit \cite{Benilov20d} shows that the ridge
dynamics depends on how the vapor-to-liquid density ratio $\rho_{v}/\rho_{l}$
compares to $\varepsilon$:

\begin{itemize}
\item If $\rho_{v}/\rho_{l}\gtrsim\varepsilon^{4/3}$, one can derive a
thin-film approximation of the DIM and identify the terms in it that disallow
the ridge solutions.

\item If $\rho_{v}/\rho_{l}\ll\varepsilon^{4/3}$, these terms vanish, and the
resulting thin-film version of the DIM coincides with that of the
Navier--Stokes equations (both admit ridge solutions).\smallskip
\end{itemize}

Note that, for common fluids at room temperature, $\rho_{v}/\rho_{l}$ is
indeed very small: for water at $T=20^{\circ}\mathrm{C}$, for example,
$\rho_{v}/\rho_{l}\approx1.7\times10^{-5}$. Thus, one can conjecture that the
observed ridges are \emph{quasi-static}: they do evolve, but their evolution
is too slow to be distinguishable from evaporation. It still remains to find
out how exactly they evolve, which seems impossible to predict using
qualitative arguments.

Besides, water on a table is surrounded by \emph{air}, not by \emph{water
vapor} as in the present formulation. It is not obvious that Theorem 1 can be
generalized for a mixture of fluids -- especially, if the temperature happens
to be supercritical for some of the components (the critical point of nitrogen
is $-147^{\circ}\mathrm{C}$, and that of oxygen is $-119^{\circ}\mathrm{C}$).

Interestingly, Theorems 1-2 cannot be extended (at least, not in a simple way)
to axisymmetric sessile drops, described by $\rho(r,z)$, where $r$ is the
polar radius. To understand why, consider the axisymmetric equivalent of the
(two-dimensional) identity (\ref{17}),%
\begin{equation}
F[\rho(0,z)]-\int_{0}^{\infty}\int_{0}^{\infty}\frac{1}{r}\left(
\frac{\partial\rho}{\partial r}\right)  ^{2}\mathrm{d}z\,\mathrm{d}%
r=F[\bar{\rho}(z)].\label{39}%
\end{equation}
Comparing (\ref{34}) to (\ref{17}), one can see that the extra term in the
former eliminates the contradiction with the fact that the right-hand side of
(\ref{39}) is always smaller than the first term on its left-hand side.

There is another setting for which Theorems 1-2 cannot be easily generalized:
if a vertical wall is present.

\bibliography{.././../bib/refs}

\begin{thebibliography}{18}%
\makeatletter
\providecommand \@ifxundefined [1]{%
 \@ifx{#1\undefined}
}%
\providecommand \@ifnum [1]{%
 \ifnum #1\expandafter \@firstoftwo
 \else \expandafter \@secondoftwo
 \fi
}%
\providecommand \@ifx [1]{%
 \ifx #1\expandafter \@firstoftwo
 \else \expandafter \@secondoftwo
 \fi
}%
\providecommand \natexlab [1]{#1}%
\providecommand \enquote  [1]{``#1''}%
\providecommand \bibnamefont  [1]{#1}%
\providecommand \bibfnamefont [1]{#1}%
\providecommand \citenamefont [1]{#1}%
\providecommand \href@noop [0]{\@secondoftwo}%
\providecommand \href [0]{\begingroup \@sanitize@url \@href}%
\providecommand \@href[1]{\@@startlink{#1}\@@href}%
\providecommand \@@href[1]{\endgroup#1\@@endlink}%
\providecommand \@sanitize@url [0]{\catcode `\\12\catcode `\$12\catcode
  `\&12\catcode `\#12\catcode `\^12\catcode `\_12\catcode `\%12\relax}%
\providecommand \@@startlink[1]{}%
\providecommand \@@endlink[0]{}%
\providecommand \url  [0]{\begingroup\@sanitize@url \@url }%
\providecommand \@url [1]{\endgroup\@href {#1}{\urlprefix }}%
\providecommand \urlprefix  [0]{URL }%
\providecommand \Eprint [0]{\href }%
\providecommand \doibase [0]{https://doi.org/}%
\providecommand \selectlanguage [0]{\@gobble}%
\providecommand \bibinfo  [0]{\@secondoftwo}%
\providecommand \bibfield  [0]{\@secondoftwo}%
\providecommand \translation [1]{[#1]}%
\providecommand \BibitemOpen [0]{}%
\providecommand \bibitemStop [0]{}%
\providecommand \bibitemNoStop [0]{.\EOS\space}%
\providecommand \EOS [0]{\spacefactor3000\relax}%
\providecommand \BibitemShut  [1]{\csname bibitem#1\endcsname}%
\let\auto@bib@innerbib\@empty
\bibitem [{\citenamefont {Hohenberg}\ and\ \citenamefont
  {Halperin}(1977)}]{HohenbergHalperin77}%
  \BibitemOpen
  \bibfield  {author} {\bibinfo {author} {\bibfnamefont {P.~C.}\ \bibnamefont
  {Hohenberg}}\ and\ \bibinfo {author} {\bibfnamefont {B.~I.}\ \bibnamefont
  {Halperin}},\ }\bibfield  {title} {\bibinfo {title} {Theory of dynamic
  critical phenomena},\ }\href {https://doi.org/10.1103/revmodphys.49.435}
  {\bibfield  {journal} {\bibinfo  {journal} {Rev. Mod. Phys.}\ }\textbf
  {\bibinfo {volume} {49}},\ \bibinfo {pages} {435} (\bibinfo {year}
  {1977})}\BibitemShut {NoStop}%
\bibitem [{\citenamefont {Gouin}(1987)}]{Gouin87}%
  \BibitemOpen
  \bibfield  {author} {\bibinfo {author} {\bibfnamefont {H.}~\bibnamefont
  {Gouin}},\ }\bibfield  {title} {\bibinfo {title} {Utilization of the {S}econd
  {G}radient {T}heory in continuum mechanics to study the motion and
  thermodynamics of liquid--vapor interfaces},\ }in\ \href
  {https://doi.org/10.1007/978-1-4613-0707-5_47} {\emph {\bibinfo {booktitle}
  {Physicochemical Hydrodynamics}}},\ \bibinfo {series} {{NATO} {ASI} Series},
  Vol.\ \bibinfo {volume} {174},\ \bibinfo {editor} {edited by\ \bibinfo
  {editor} {\bibfnamefont {M.~G.}\ \bibnamefont {Velarde}}}\ (\bibinfo
  {publisher} {Springer {US}},\ \bibinfo {year} {1987})\ pp.\ \bibinfo {pages}
  {667--682}\BibitemShut {NoStop}%
\bibitem [{\citenamefont {Anderson}\ \emph {et~al.}(1998)\citenamefont
  {Anderson}, \citenamefont {McFadden},\ and\ \citenamefont
  {Wheeler}}]{AndersonMcFaddenWheeler98}%
  \BibitemOpen
  \bibfield  {author} {\bibinfo {author} {\bibfnamefont {D.~M.}\ \bibnamefont
  {Anderson}}, \bibinfo {author} {\bibfnamefont {G.~B.}\ \bibnamefont
  {McFadden}},\ and\ \bibinfo {author} {\bibfnamefont {A.~A.}\ \bibnamefont
  {Wheeler}},\ }\bibfield  {title} {\bibinfo {title} {Diffuse-interface methods
  in fluid mechanics},\ }\href {https://doi.org/10.1146/annurev.fluid.30.1.139}
  {\bibfield  {journal} {\bibinfo  {journal} {Annu. Rev. Fluid Mech.}\ }\textbf
  {\bibinfo {volume} {30}},\ \bibinfo {pages} {139} (\bibinfo {year}
  {1998})}\BibitemShut {NoStop}%
\bibitem [{\citenamefont {Pismen}\ and\ \citenamefont
  {Pomeau}(2000)}]{PismenPomeau00}%
  \BibitemOpen
  \bibfield  {author} {\bibinfo {author} {\bibfnamefont {L.~M.}\ \bibnamefont
  {Pismen}}\ and\ \bibinfo {author} {\bibfnamefont {Y.}~\bibnamefont
  {Pomeau}},\ }\bibfield  {title} {\bibinfo {title} {Disjoining potential and
  spreading of thin liquid layers in the diffuse-interface model coupled to
  hydrodynamics},\ }\href {https://doi.org/10.1103/physreve.62.2480} {\bibfield
   {journal} {\bibinfo  {journal} {Phys. Rev. E}\ }\textbf {\bibinfo {volume}
  {62}},\ \bibinfo {pages} {2480} (\bibinfo {year} {2000})}\BibitemShut
  {NoStop}%
\bibitem [{\citenamefont {Jacqmin}(2000)}]{Jacqmin00}%
  \BibitemOpen
  \bibfield  {author} {\bibinfo {author} {\bibfnamefont {D.}~\bibnamefont
  {Jacqmin}},\ }\bibfield  {title} {\bibinfo {title} {Contact-line dynamics of
  a diffuse fluid interface},\ }\href
  {https://doi.org/10.1017/s0022112099006874} {\bibfield  {journal} {\bibinfo
  {journal} {J. Fluid Mech.}\ }\textbf {\bibinfo {volume} {402}},\ \bibinfo
  {pages} {57} (\bibinfo {year} {2000})}\BibitemShut {NoStop}%
\bibitem [{\citenamefont {Thiele}\ \emph {et~al.}(2007)\citenamefont {Thiele},
  \citenamefont {Madruga},\ and\ \citenamefont
  {Frastia}}]{ThieleMadrugaFrastia07}%
  \BibitemOpen
  \bibfield  {author} {\bibinfo {author} {\bibfnamefont {U.}~\bibnamefont
  {Thiele}}, \bibinfo {author} {\bibfnamefont {S.}~\bibnamefont {Madruga}},\
  and\ \bibinfo {author} {\bibfnamefont {L.}~\bibnamefont {Frastia}},\
  }\bibfield  {title} {\bibinfo {title} {Decomposition driven interface
  evolution for layers of binary mixtures. {I}. {M}odel derivation and
  stratified base states},\ }\href {https://doi.org/10.1063/1.2824404}
  {\bibfield  {journal} {\bibinfo  {journal} {Phys. Fluids}\ }\textbf {\bibinfo
  {volume} {19}},\ \bibinfo {pages} {122106} (\bibinfo {year}
  {2007})}\BibitemShut {NoStop}%
\bibitem [{\citenamefont {Ding}\ and\ \citenamefont
  {Spelt}(2007)}]{DingSpelt07}%
  \BibitemOpen
  \bibfield  {author} {\bibinfo {author} {\bibfnamefont {H.}~\bibnamefont
  {Ding}}\ and\ \bibinfo {author} {\bibfnamefont {P.~D.~M.}\ \bibnamefont
  {Spelt}},\ }\bibfield  {title} {\bibinfo {title} {Wetting condition in
  diffuse interface simulations of contact line motion},\ }\href
  {https://doi.org/10.1103/physreve.75.046708} {\bibfield  {journal} {\bibinfo
  {journal} {Phys. Rev. E}\ }\textbf {\bibinfo {volume} {75}},\ \bibinfo
  {pages} {046708} (\bibinfo {year} {2007})}\BibitemShut {NoStop}%
\bibitem [{\citenamefont {Yue}\ and\ \citenamefont {Feng}(2011)}]{YueFeng11}%
  \BibitemOpen
  \bibfield  {author} {\bibinfo {author} {\bibfnamefont {P.}~\bibnamefont
  {Yue}}\ and\ \bibinfo {author} {\bibfnamefont {J.~J.}\ \bibnamefont {Feng}},\
  }\bibfield  {title} {\bibinfo {title} {Can diffuse-interface models
  quantitatively describe moving contact lines?},\ }\href
  {https://doi.org/10.1140/epjst/e2011-01434-y} {\bibfield  {journal} {\bibinfo
   {journal} {Eur. Phys. J. Spec. Top.}\ }\textbf {\bibinfo {volume} {197}},\
  \bibinfo {pages} {37} (\bibinfo {year} {2011})}\BibitemShut {NoStop}%
\bibitem [{\citenamefont {Sibley}\ \emph {et~al.}(2013)\citenamefont {Sibley},
  \citenamefont {Nold}, \citenamefont {Savva},\ and\ \citenamefont
  {Kalliadasis}}]{SibleyNoldSavvaKalliadasis13b}%
  \BibitemOpen
  \bibfield  {author} {\bibinfo {author} {\bibfnamefont {D.~N.}\ \bibnamefont
  {Sibley}}, \bibinfo {author} {\bibfnamefont {A.}~\bibnamefont {Nold}},
  \bibinfo {author} {\bibfnamefont {N.}~\bibnamefont {Savva}},\ and\ \bibinfo
  {author} {\bibfnamefont {S.}~\bibnamefont {Kalliadasis}},\ }\bibfield
  {title} {\bibinfo {title} {The contact line behaviour of solid-liquid-gas
  diffuse-interface models},\ }\href {https://doi.org/10.1063/1.4821288}
  {\bibfield  {journal} {\bibinfo  {journal} {Phys. Fluids}\ }\textbf {\bibinfo
  {volume} {25}},\ \bibinfo {pages} {092111} (\bibinfo {year}
  {2013})}\BibitemShut {NoStop}%
\bibitem [{\citenamefont {Kusumaatmaja}\ \emph {et~al.}(2016)\citenamefont
  {Kusumaatmaja}, \citenamefont {Hemingway},\ and\ \citenamefont
  {Fielding}}]{KusumaatmajaHemingwayFielding16}%
  \BibitemOpen
  \bibfield  {author} {\bibinfo {author} {\bibfnamefont {H.}~\bibnamefont
  {Kusumaatmaja}}, \bibinfo {author} {\bibfnamefont {E.~J.}\ \bibnamefont
  {Hemingway}},\ and\ \bibinfo {author} {\bibfnamefont {S.~M.}\ \bibnamefont
  {Fielding}},\ }\bibfield  {title} {\bibinfo {title} {Moving contact line
  dynamics: from diuse to sharp interfaces},\ }\href
  {https://doi.org/10.1017/jfm.2015.697} {\bibfield  {journal} {\bibinfo
  {journal} {J. Fluid Mech.}\ }\textbf {\bibinfo {volume} {788}},\ \bibinfo
  {pages} {209} (\bibinfo {year} {2016})}\BibitemShut {NoStop}%
\bibitem [{\citenamefont {Fakhari}\ and\ \citenamefont
  {Bolster}(2017)}]{FakhariBolster17}%
  \BibitemOpen
  \bibfield  {author} {\bibinfo {author} {\bibfnamefont {A.}~\bibnamefont
  {Fakhari}}\ and\ \bibinfo {author} {\bibfnamefont {D.}~\bibnamefont
  {Bolster}},\ }\bibfield  {title} {\bibinfo {title} {Diffuse interface
  modeling of three-phase contact line dynamics on curved boundaries: {A}
  lattice {B}oltzmann model for large density and viscosity ratios},\ }\href
  {https://doi.org/10.1016/j.jcp.2017.01.025} {\bibfield  {journal} {\bibinfo
  {journal} {J. Comput. Phys.}\ }\textbf {\bibinfo {volume} {334}},\ \bibinfo
  {pages} {620} (\bibinfo {year} {2017})}\BibitemShut {NoStop}%
\bibitem [{\citenamefont {Borcia}\ \emph {et~al.}(2019)\citenamefont {Borcia},
  \citenamefont {Borcia}, \citenamefont {Bestehorn}, \citenamefont {Varlamova},
  \citenamefont {Hoefner},\ and\ \citenamefont
  {Reif}}]{BorciaBorciaBestehornVarlamovaHoefnerReif19}%
  \BibitemOpen
  \bibfield  {author} {\bibinfo {author} {\bibfnamefont {R.}~\bibnamefont
  {Borcia}}, \bibinfo {author} {\bibfnamefont {I.~D.}\ \bibnamefont {Borcia}},
  \bibinfo {author} {\bibfnamefont {M.}~\bibnamefont {Bestehorn}}, \bibinfo
  {author} {\bibfnamefont {O.}~\bibnamefont {Varlamova}}, \bibinfo {author}
  {\bibfnamefont {K.}~\bibnamefont {Hoefner}},\ and\ \bibinfo {author}
  {\bibfnamefont {J.}~\bibnamefont {Reif}},\ }\bibfield  {title} {\bibinfo
  {title} {Drop behavior influenced by the correlation length on noisy
  surfaces},\ }\href {https://doi.org/10.1021/acs.langmuir.8b03878} {\bibfield
  {journal} {\bibinfo  {journal} {Langmuir}\ }\textbf {\bibinfo {volume}
  {35}},\ \bibinfo {pages} {928} (\bibinfo {year} {2019})}\BibitemShut
  {NoStop}%
\bibitem [{\citenamefont {Gallo}\ \emph {et~al.}(2020)\citenamefont {Gallo},
  \citenamefont {Magaletti}, \citenamefont {Cocco},\ and\ \citenamefont
  {Casciola}}]{GalloMagalettiCoccoCasciola20}%
  \BibitemOpen
  \bibfield  {author} {\bibinfo {author} {\bibfnamefont {M.}~\bibnamefont
  {Gallo}}, \bibinfo {author} {\bibfnamefont {F.}~\bibnamefont {Magaletti}},
  \bibinfo {author} {\bibfnamefont {D.}~\bibnamefont {Cocco}},\ and\ \bibinfo
  {author} {\bibfnamefont {C.~M.}\ \bibnamefont {Casciola}},\ }\bibfield
  {title} {\bibinfo {title} {Nucleation and growth dynamics of vapour
  bubbles},\ }\href {https://doi.org/10.1017/jfm.2019.844} {\bibfield
  {journal} {\bibinfo  {journal} {J. Fluid Mech.}\ }\textbf {\bibinfo {volume}
  {883}},\ \bibinfo {pages} {A14} (\bibinfo {year} {2020})}\BibitemShut
  {NoStop}%
\bibitem [{\citenamefont {Gelissen}\ \emph {et~al.}(2020)\citenamefont
  {Gelissen}, \citenamefont {van~der Geld}, \citenamefont {Baltussen},\ and\
  \citenamefont {Kuerten}}]{GelissenVandergeldBaltussenKuerten20}%
  \BibitemOpen
  \bibfield  {author} {\bibinfo {author} {\bibfnamefont {E.~J.}\ \bibnamefont
  {Gelissen}}, \bibinfo {author} {\bibfnamefont {C.~W.~M.}\ \bibnamefont
  {van~der Geld}}, \bibinfo {author} {\bibfnamefont {M.~W.}\ \bibnamefont
  {Baltussen}},\ and\ \bibinfo {author} {\bibfnamefont {J.~G.~M.}\ \bibnamefont
  {Kuerten}},\ }\bibfield  {title} {\bibinfo {title} {Modeling of droplet
  impact on a heated solid surface with a diffuse interface model},\ }\href
  {https://doi.org/10.1016/j.ijmultiphaseflow.2019.103173} {\bibfield
  {journal} {\bibinfo  {journal} {Int. J. Multiphase Flow}\ }\textbf {\bibinfo
  {volume} {123}},\ \bibinfo {pages} {103173} (\bibinfo {year}
  {2020})}\BibitemShut {NoStop}%
\bibitem [{\citenamefont {Benilov}(2020{\natexlab{a}})}]{Benilov20a}%
  \BibitemOpen
  \bibfield  {author} {\bibinfo {author} {\bibfnamefont {E.~S.}\ \bibnamefont
  {Benilov}},\ }\bibfield  {title} {\bibinfo {title} {The dependence of the
  surface tension and contact angle on the temperature, as described by the
  diffuse-interface model},\ }\bibfield  {journal} {\bibinfo  {journal} {Phys.
  Rev. E}\ }\href {https://doi.org/10.1103/PhysRevE.101.042803}
  {10.1103/PhysRevE.101.042803} (\bibinfo {year}
  {2020}{\natexlab{a}})\BibitemShut {NoStop}%
\bibitem [{\citenamefont {Benilov}(2020{\natexlab{b}})}]{Benilov20b}%
  \BibitemOpen
  \bibfield  {author} {\bibinfo {author} {\bibfnamefont {E.~S.}\ \bibnamefont
  {Benilov}},\ }\bibfield  {title} {\bibinfo {title} {Asymptotic reductions of
  the diffuse-interface model, with applications to contact lines in fluids},\
  }\href@noop {} {\bibfield  {journal} {\bibinfo  {journal} {to appear in Phys.
  Rev. Fluids (also available on \url{https://arxiv.org/abs/1907.04434})}\ }
  (\bibinfo {year} {2020}{\natexlab{b}})}\BibitemShut {NoStop}%
\bibitem [{\citenamefont {Courant}\ and\ \citenamefont
  {Hilbert}(1989)}]{CourantHilbert89}%
  \BibitemOpen
  \bibfield  {author} {\bibinfo {author} {\bibfnamefont {R.}~\bibnamefont
  {Courant}}\ and\ \bibinfo {author} {\bibfnamefont {D.}~\bibnamefont
  {Hilbert}},\ }\href {https://doi.org/10.1002/9783527617210} {\emph {\bibinfo
  {title} {Methods of Mathematical Physics}}}\ (\bibinfo  {publisher} {Wiley},\
  \bibinfo {year} {1989})\BibitemShut {NoStop}%
\bibitem [{\citenamefont {Benilov}(2020{\natexlab{c}})}]{Benilov20d}%
  \BibitemOpen
  \bibfield  {author} {\bibinfo {author} {\bibfnamefont {E.~S.}\ \bibnamefont
  {Benilov}},\ }\bibfield  {title} {\bibinfo {title} {Dynamics of liquid films,
  as described by the diffuse-interface model},\ }\href@noop {} {\bibfield
  {journal} {\bibinfo  {journal} {\url{https://arxiv.org/abs/2007.03351}}\ }
  (\bibinfo {year} {2020}{\natexlab{c}})}\BibitemShut {NoStop}%
\end{thebibliography}%

\end{document}